\documentclass[twocolumn,nofootinbib,showpacs]{revtex4}

\usepackage{graphicx}
\usepackage{dcolumn}
\usepackage{amsmath,amssymb,epsfig}
\usepackage{paralist}
\usepackage{comment}
\usepackage{graphicx}
\usepackage{multirow}

\begin{document}

\title{Support for temporally varying behavior of the Pioneer anomaly \\
from the extended Pioneer 10 and 11 Doppler data sets}

\author{Slava G. Turyshev$^1$,
Viktor T. Toth$^2$, Jordan Ellis$^1$, and Craig B. Markwardt$^3$
}

\affiliation{\vskip 3pt
$^1$Jet Propulsion Laboratory, California Institute of Technology,\\
4800 Oak Grove Drive, Pasadena, CA 91109-0899, USA
}%

\affiliation{\vskip 3pt
$^2$Ottawa, ON, Canada,
}%

\affiliation{\vskip 3pt
$^3$NASA Goddard Space Flight Center, MD, USA
}%

\date{\today}

\begin{abstract}
The Pioneer anomaly is a small sunward anomalous acceleration found in the trajectory analysis of the Pioneer 10 and 11 spacecraft. As part of the investigation of the effect, analysis of recently recovered Doppler data for both spacecraft has been completed. The presence of a small anomalous acceleration is confirmed using data spans more than twice as long as those that were previously analyzed. We examine the constancy and direction of the Pioneer anomaly, and conclude that:
\begin{inparaenum}[i)]
\item the data favor a temporally decaying anomalous acceleration ($\sim2\times 10^{-11}$ m/s$^2$/yr) with an over 10\% improvement in the residuals compared to a constant acceleration model;
\item although the direction of the acceleration remains imprecisely determined, we find no support in favor of a Sun-pointing direction over the Earth-pointing or along the spin-axis directions, and
\item support for an early ``onset'' of the acceleration remains weak in the pre-Saturn Pioneer 11 tracking data.
\end{inparaenum}
We present these new findings and discuss their implications for the nature of the Pioneer anomaly.
\end{abstract}


\maketitle

\section{Introduction}

Analysis of the navigational tracking data received from the Pioneer 10 and 11 spacecraft at large heliocentric distances of $\sim$20--70 AU indicated the presence of a small anomalous Doppler frequency drift in their radio-metric observables \cite{pioprl,moriond}. Ultimately, this drift was  interpreted as an anomalous constant acceleration acting on both of these spacecraft in the sunward direction, with a magnitude of $a_P=(8.74\pm 1.33)\times 10^{-10}$~m/s$^2$ \cite{pioprd}. There were also earlier reports that, in the case of Pioneer 11, the anomaly may have begun with a relatively sudden ``onset'' \cite{pioprd} shortly after the spacecraft's encounter with Saturn. This acceleration of unknown origin is today known as the Pioneer anomaly (for a review, see \cite{piolrr}). In this paper, we present an analysis of newly recovered Doppler data for both Pioneer 10 and 11, and probe the question of the constancy of the anomalous acceleration, as well as place constraints on its direction.

\section{Data sets and Analysis}

Following the 2002 study \cite{pioprd}, an effort was initiated to collect all available Doppler data for both spacecraft \cite{MDR2005,piolrr}.  Using standard data conditioning techniques \cite{pioprd}, several decades-old archival data tracking files (see discussion in \cite{MDR2005}) were converted to modern formats, processed, edited and filtered for corrupt data.  Uplink frequency records at NASA's Deep Space Network (DSN) were reconstructed from redundant information. We summarize the newly recovered Pioneer 10 \& 11 Doppler data sets in Table \ref{tab:datasets}, including the range of dates covered and number of two-way and three-way \cite{pioprd,piolrr} coherent Doppler data points available.  The largest portion of the data lies in ``deep space'' (``DS'') beyond planetary encounters (heliocentric distances of 18--80 AU and 9--32 AU for Pioneer 10 \& 11 respectively).  The data also include planetary encounters with Jupiter (``J'') and Saturn (``S'', for Pioneer 11 only).  Data for other time ranges did not survive to the present.  Here, we focus on the deep space data, and the portion of Pioneer 11 Saturn data which is primarily under the gravitational influence of the Sun, which we designate Saturn ``approach'' (``SA'').  These latter data are outside of Saturn's Roche (or Hill) radius so that Saturn and its moons can be treated only as a perturbing influence.

Compared to the 2002 data, the data arc for Pioneer 10 has doubled in length, from 11.5~years to 23.1 years, and the total number of data points increased from 20,055 to 41,054, including Jupiter encounter. For Pioneer 11, the length of the contiguous data arc increased from 3.75 to 10.75 years, and the number of data points increased from 10,616 to 81,537, including encounters. In terms of heliocentric distance, the Pioneer 10 \& 11 arcs overlap from 18--32 AU.

\begin{table}
\caption{Pioneer Doppler Data Sets
\label{tab:datasets}}
\begin{ruledtabular}
\begin{tabular}{lccrr}
S/C\footnotemark[1]
    & Data\footnotemark[2]
         & Date Range                     & Points & Man.\footnotemark[3]\\
\hline
P10 & J  & 1973-10-15 -- 1973-12-27 &  5806  & $>$6 \\
    & DS & 1979-02-14 -- 2002-03-03 & 35248  & $>$83\\
    & 2002DS & 1987-01-03 -- 1998-07-22 & 20055 & 28 \\

\hline
P11 & J  & 1974-04-18  -- 1974-12-26  &  7467  & $>$16\\
    & S  & 1977-11-01  -- 1979-09-18  &  9017  & $>$35\\
    & SA & 1977-11-01  -- 1979-06-29  &  4282  & ---  \\
    & DS & 1980-01-12  -- 1990-10-01  & 65053  & $>$92\\
    & 2002DS & 1987-01-05 -- 1990-10-01 & 10616 & 22 \\
\end{tabular}
\end{ruledtabular}
\footnotetext[1]{(P10), (P11) denote Pioneer 10 and Pioneer 11 respectively.}
\footnotetext[2]{(J) Jupiter; (S) Saturn; and (SA, subset of S) Saturn approach; (DS) deep space; (2002DS) is the data used in the 2002 study  \cite{pioprd}.}
\footnotetext[3]{Number of attitude maneuvers performed.}
\end{table}

During the mission, both spacecraft performed numerous maneuvers.  As the Pioneers were spin-stabilized, periodic attitude maneuvers were required to realign the spin and antenna axis to the spacecraft-Earth direction to within 1.5$^\circ$.  While attitude maneuvers used a fore-aft pair of thrusters in tandem, small residual impulses along the spin direction may occur for \cite{piolrr}.  Information about maneuvers is available from recovered mission records and from the spacecraft telemetry \cite{MDR2005}, and is summarized in Table \ref{tab:datasets}. The data given are lower limits as it is possible that a few maneuvers are not reflected in the available telemetry.

Our analysis was carried out using the Jet Propulsion Laboratory's (JPL) Orbit Determination Program (ODP) that was used for earlier work on the anomaly \cite{pioprd,piolrr}. As before, central to the analysis was establishing a model orbit for the spacecraft that takes into account all known forces, gravitational and non-gravitational, while numerically integrating the appropriate equations of motion. The model included the effects of planetary perturbations, solar radiation pressure, propulsive maneuvers, general relativity, and bias and drift in the Doppler observable. Planetary coordinates and the solar system masses were obtained using JPL's Export Planetary Ephemeris DE421. The model also included the precise positions of Earth-based stations of the DSN as well as radio propagation effects (see details in \cite{piolrr}).

The study of the deep space data arcs (DS, see Table~\ref{tab:datasets}) from the extended Doppler data set was expected to improve our understanding of two key characteristics of the Pioneer anomaly: its direction and its temporal behavior. Additionally, it was expected that some of the early data, in particular the Pioneer 11 Saturn approach data (SA), would help us confirm whether or not the anomalous behavior began with a relatively sudden ``onset'' \cite{pioprd}.

We considered three models for the anomalous acceleration --- constant, linear and exponential --- all applied along the nominal Earth-spacecraft line. The constant model has one parameter, $a_P$, representing a constant modeling error.  The linear model,
\begin{equation}
a_P(t) = a_P(t_0) + (t-t_0)\dot{a}_P \label{eq:jerk}
\end{equation}
contains a jerk term, $\dot{a}_P$.  The exponential model,
\begin{equation}
a_P(t) = a_P(t_0) e^{-\beta(t-t_0)\ln{2}} \label{eq:exp}
\end{equation}
decays with half life $\beta^{-1}$.  This last model is physically motivated by a potential relation to the on-board power generators, which radioactively decay. The epoch is $t_0=$January 1, 1972.

Separately, the anomalous acceleration was estimated using a batched stochastic model \cite{pioprd,piolrr}. This method produces a smoothed acceleration for each batch \cite{pioprd}.  As the model with the most estimated parameters, it is likely to produce the best possible fit, but since it is purely phenomenological, it provides the least physical insight into the anomaly.

To consider the direction of the anomaly, we separately modeled the acceleration as a constant vector in four principal directions \cite{pioprd,piolrr}: that of i) the Earth, ii) the Sun, iii) the spin axis, and iv) the spacecraft velocity vector. Of these, the spin and Earth-spacecraft axes are effectively degenerate as the spacecraft were maintaining an Earth orientation for continuous radio communication. We estimated an acceleration vector that was constant in a reference frame with its $z$-axis aligned with the spacecraft-Sun line, and an acceleration vector that was constant in a reference frame with its $z$-axis aligned along the spacecraft-Earth line. The $x$-axis in both cases lay in the plane of the ecliptic, and the $y$-axis completed a right-handed triad. ODP solves for the components of the acceleration vector independently.

For all acceleration models, the initial state vector and the velocity impulses for each attitude maneuver were also estimated using a least squares fit.

The estimate and formal errors for the model parameters are computed by the least squares fit assuming the Doppler measurement errors are uncorrelated. In reality, the residual still shows significant structure, perhaps due to mismodeling (e.g., models of solar plasma, the atmosphere, etc.). The presence of such autocorrelation in the residuals is the reason why it is common for ``realistic'' errors to be larger than formal errors by an order of magnitude or more. In this Letter we report formal errors, with the understanding that these values do not necessarily represent the actual uncertainty of model parameters, but can be used as an indication of the overall goodness of the fit and thus, the quality of the model.

Acceleration models are compared computing $\sigma_\nu$, the root mean square (RMS) of the difference between modeled and observed Doppler frequencies. A typical ``good'' fit should yield RMS residuals of 10 mHz or less, often under 5 mHz. An RMS residual above $\sim$10 mHz usually indicates that the model does not adequately describe the physics of the spacecraft's motion, the dynamics of the solar system, or the propagation of the radio signal. We find that the Pioneer 10 data set is noisier than that for Pioneer 11, with RMS residuals differing by a factor of two.

We evaluated the models described in the previous section using the ODP.  The deep space data sets of both the Pioneer 10 and 11
spacecraft\footnote{For technical reasons, Pioneer 10 data up to 1998 and Pioneer 11 data for 1983--1990 were used.} were evaluated using the one-dimensional (1-D) constant, linear, and exponential models, and the results are shown in Table \ref{tab:deepspace}. The fit quality of the different temporal models is nearly the same.  The variable models are consistent with a gradually decreasing acceleration, either at a rate of $\sim1.7\times 10^{-11}$ m/s$^2$/yr, or a half life of $\sim$27 yr averaged for both spacecraft.

\begin{figure}[t!]
\includegraphics[width=\linewidth]{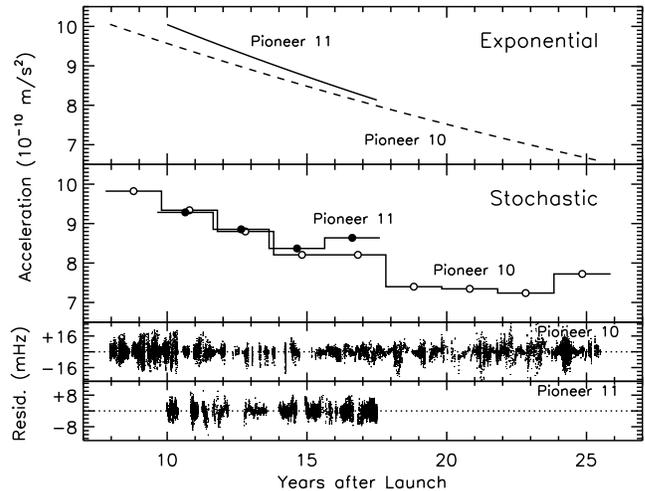}
\vskip -10pt
\caption{\label{fig:stochastic} {\it Top panel}: Estimates of the anomalous acceleration of Pioneer 10 (dashed line) and Pioneer 11 (solid line) using an exponential model. {\it Second panel}: Stochastic acceleration estimates for Pioneer 10 (open circles) and Pioneer 11 (filled circles), shown as step functions. {\it Bottom two panels}: Doppler residuals of the stochastic acceleration model. Note the difference in vertical scale for Pioneer 10 vs. Pioneer 11.
}

\end{figure}

The results of the exponential model are shown in Figure~\ref{fig:stochastic}. The figure also shows the stochastic estimate, illustrating the similarity between the parameterized and stochastic models. The RMS residuals of the stochastic fit show further improvement compared to other models:
\begin{equation}
\sigma_\mathrm{P10}=4.40~\mathrm{mHz},\qquad
\sigma_\mathrm{P11}=2.02~\mathrm{mHz}.
\end{equation}

The results of the three-dimensional (3-D) models are shown in Table \ref{tab:vectparam}. For the DS data sets, acceleration vectors that are constant in a solar system barycentric reference frame or constant relative to the spacecraft-Earth line cannot be distinguished. The estimated vector direction is within 6$^\circ$ of the spacecraft spin axis.

We also investigated the possibility of an ``onset'' of the anomaly using Pioneer 11 Saturn Approach (``SA'') data from 1977--1979, the results of which are also shown in Table \ref{tab:vectparam}.  While we attempted to fit a simple 1-D Earth pointing constant acceleration, results were poor\footnote{The validity of the SA results was questioned because of the large uncertainty in the acceleration estimates, their sensitivity to solar radiation pressure parameters, the structure of the residual and presence of a 21~m/s maneuver. In the vicinity of this maneuver, the residual display shows residuals of 20~mHz, five times greater than the typical residuals outside this region.}. The vector model has very large errors in the $x$ direction, which relate to a larger solar radiation pressure effect at these distances.  The acceleration magnitude for Saturn approach is $a_\mathrm{P11}=(4.58\pm 11.80)\times 10^{-10}$~m/s$^2$. This result is consistent with the acceleration estimates of the DS phase, and thus, we cannot conclude definitively that an ``onset'' exists or not (see \cite{pioprd,piolrr} for discussion).

\begin{table}[t]
\caption{Deep Space Acceleration Parameters
\label{tab:deepspace}}
\begin{ruledtabular}
\begin{tabular}{llccc}
S/C\footnotemark[1]
    & Model        & $\sigma_\nu$
                           & $a_P$\footnotemark[2]
                                      & Additional\\
    &              &   mHz &  $10^{-10}$ m/s$^2$ &  parameter\footnotemark[2]$^,$\footnotemark[3]\\
\hline
P10 & Constant     & 4.98  &  8.17(2) &           \\
    & Linear       & 4.60  & 11.06(8)~ & $\dot{a}_P$ = $-$0.17(1) \\
    & Exponential  & 4.58  & \,12.22(16)& $\beta^{-1}$ = 28.8(0.7) \\
\hline
P11 & Constant     & 3.67  &  9.15(7) &           \\
    & Linear       & 2.09  & 11.65(42)& $\dot{a}_P$ = $-$0.18(3) \\
    & Exponential  & 2.06  & 13.79(62)& $\beta^{-1}$ = 24.6(2.4)\\
\end{tabular}
\end{ruledtabular}
\footnotetext[1]{See Table \ref{tab:datasets} for designations.}
\footnotetext[2]{1-$\sigma$ formal error quoted in final digit(s).}
\footnotetext[3]{$\dot{a}_P$ in $10^{-10}$ m/s$^2$/yr, $\beta^{-1}$ in yr.}
\end{table}

\begin{table}
\caption{Vector Acceleration Parameters
\label{tab:vectparam}}
\begin{ruledtabular}
\begin{tabular}{lllcccc}
S/C\footnotemark[1]
    & Data
           & Center  & $\sigma_\nu$
                              & $a_{P,z}$\footnotemark[2]& $a_{P,x}$\footnotemark[2]  & $a_{P,y}$\footnotemark[2]  \\
    &      &         &   mHz &  \multicolumn{3}{c}{$10^{-10}$ m/s$^2$\ \ \ }  \\
\hline
P10 & DS   & Earth   &   4.98 & 8.17(2)  &  ~~0.42(3)   & 0.08(1)    \\
    &      & Sun     &   4.98 & 8.17(2)  &  ~~0.42(3)   & 0.08(1)    \\
\hline
P11 & DS   &  Earth  &   2.13 & 8.62(6)  & $-$0.73(3) & 0.97(1)    \\
    &      &  Sun    &   2.21 & 8.64(6)  & $-$0.76(3) & 0.97(1)    \\
\hline
P11 & SA   & Earth/   &   3.61 & 4.48     & $-$1.38    &$-$0.27       \\
    &      & Sun \footnotemark[3]
                      &        & $\pm$0.50& $\pm$11.33 & $\pm$1.7  \\
\end{tabular}
\end{ruledtabular}
\footnotetext[1]{See Table \ref{tab:datasets} for designations.}
\footnotetext[2]{1-$\sigma$ error quoted in final digit, except as noted.}
\footnotetext[3]{Results for Earth-/Sun-pointing were similar.}
\end{table}

\section{Discussion}

{\it Temporal behavior:} We can unambiguously confirm the presence of an anomalous acceleration the recently recovered Pioneer 10 \& 11 spacecraft tracking data, with consistent magnitudes between the two spacecraft, and also consistent with previous results \cite{pioprl,pioprd,CBM2005,TOTH2008}.

The constant acceleration models, both 1-D and 3-D, are essentially identical for Pioneer 10.  However, for Pioneer 11, the RMS residuals improve when considering an unknown constant force, perpendicular to the spacecraft-Earth direction. This may be related to the anomalous spin-up of Pioneer 11 \cite{MDR2005}.

Our estimates of a jerk term (\ref{eq:jerk}) are consistent with earlier studies \cite{piolrr}. Markwardt \cite{CBM2005} obtained an improved fit of Pioneer 10 data when estimating a jerk of $\dot{a}_\mathrm{P10}=-0.18\times 10^{-10}$~m/s$^2$/yr; also Toth \cite{TOTH2008} obtained $\dot{a}_\mathrm{P10}=(-0.21\pm 0.04)\times 10^{-10}$~m/s$^2$/yr, $\dot{a}_\mathrm{P11}=(-0.34\pm 0.12)$~m/s$^2$/yr for Pioneer 10 \& 11, respectively.

The rationale for an exponential model (\ref{eq:exp}) is based on the possibility that the acceleration may be due to thermal recoil forces generated on-board. Due to degradation of the RTG thermocouples and changes in the thermal louver system \cite{piolrr}, the resulting thermal recoil force could have a half-life significantly shorter than the 87.74~year half-life of the $^{238}$Pu fuel \cite{Toth:2007}, with 27 years being in the acceptable range.

The gradually decreasing linear and exponential decay models yield marginally improved fits when compared to the constant acceleration model, as does the stochastic model. The presence of maneuvers confounds our ability to detect such terms unambiguously. The addition of earlier data arcs, with greater occurrences of maneuvers did not help as much as desired.

Our measure of goodness-of-fit, $\sigma_\nu$, may allow comparison of competing models using the standard $F$-ratio statistic (i.e. ratio-of-variances), even if $\sigma_\nu$ itself is not a standard statistic.  Preliminary results indicate that the time variable models (linear, exponential, stochastic) are indeed better than the simple constant models, for both spacecraft trajectories.  Future work will attempt to quantify this improvement more rigorously.

{\it Direction:} For continuous communication, it was necessary to orient the spacecraft so as to keep the Earth with the 3$^\circ$ of their antenna beamwidth. The Sun-probe-Earth (SPE) angle remained small and varied only from 2.6$^\circ$ (1980) to 0.7$^\circ$ (2001) for Pioneer 10 and from 6.0$^\circ$ (1980) to 1.7$^\circ$ (1990) for Pioneer 11. The solar plasma noise present in the data rendered the effort to distinguish these nearly coincident directions fruitless. There was also a possibility that Pioneer 11 data from prior to Saturn encounter, when the SPE angle was larger, would allow us to distinguish between the Earth (and spin axis) vs. Sun directions. Unfortunately, these hopes were, too, in vain because of solar plasma noise, a malfunctioning thruster, and frequent maneuvers. Although the Earth direction is marginally preferred by the solution (see Table~\ref{tab:vectparam}), the Sun, the Earth, and the spin axis directions cannot be distinguished.

We can exclude an anomaly directed along the spacecraft velocity vector.  In 1980, the angle between the best-fit acceleration vector and the spacecraft velocity vector is 8.5$^\circ$ (Pioneer 10) and 31.8$^\circ$ (Pioneer 11).  This was sufficient to show that the anomaly is not directed along the velocity vector using Toth's orbit determination program \cite{TOTH2008} with the Pioneer 11 DS data arc.

{\it Onset:} The Doppler data obtained using Pioneer 11 prior to its Saturn encounter are consistent with a possible onset, but the uncertainty remains large. Some, or all, of this onset may be due to mismodeling the effects of solar pressure. However, the relative shortness of these data arcs and the large number of maneuvers performed during this period (including a significant trajectory correction maneuver) make it difficult to reach a robust conclusion.

{\it Origin:} The most likely cause of the Pioneer anomaly is the anisotropic emission of on-board heat. This fact was recognized early on \cite{pioprd}, leading to a detailed thermal analysis of the Pioneer 10/11 spacecraft. In a parallel effort, using recovered project documentation and telemetry records, a highly detailed finite-element thermal model of the two spacecraft was constructed and used to estimate the recoil force due to anisotropically radiated on-board generated heat at various heliocentric distances  \cite{Toth:2007,piolrr}. A conclusive result can only be reached by incorporating the thermal recoil force, computed as a function of time, into the standard set of spacecraft force models that are used for Doppler analysis  \cite{Toth2009,piolrr}. Such an analysis was initiated once the extended Pioneer 10 and 11 Doppler data sets became available.  The main question is whether or not a statistically significant anomalous acceleration signal still remains in the residuals after the thermal recoil force has been properly accounted for. Results of this meticulous study will be published soon.

\begin{acknowledgments}

We thank G.L. Goltz, K.J. Lee, N.A. Mottinger of JPL for their indispensable help with the Pioneer Doppler data recovery. We thank S.W. Asmar, W.M. Folkner, T.P. McElrath, M.M. Watkins, and J.G. Williams of JPL for their interest, support and encouragement during the work and preparation of this manuscript. We also thank The Planetary Society for their continuing interest in the Pioneer anomaly and their support.
Some aspects of this work were developed at the International Space Science Institute (ISSI), Bern, Switzerland, for which ISSI's hospitality and support are kindly acknowledged.
This work in part was performed at the Jet Propulsion Laboratory, California Institute of Technology, under a contract with the National Aeronautics and Space Administration.

\end{acknowledgments}

\bibliography{new-doppler}

\end{document}